\documentclass{aa}

\usepackage[varg]{txfonts}
\usepackage{graphicx,amssymb,natbib}
\usepackage{txfonts}
\usepackage{mathrsfs}

\usepackage{color}

\usepackage[switch, modulo]{lineno}

\begin{document}

\def\A{\,\AA\ }
\def\AF{\,\AA}
\newcommand{\etal}{et~al. }
\newcommand{\ha}{${\rm H\alpha}$}
\newcommand{\ca}{\ion{Ca}{ii} 8542\A}
\newcommand{\calc}{\ion{Ca}{ii}}
\newcommand{\fca}{\ion{Ca}{ii} 8542~\AA}
\newcommand{\cmv}{\mbox{{\rm\thinspace cm$^{-3}$}}}
\newcommand{\cmmf}{\mbox{{\rm\thinspace cm$^{-5}$}} }
\newcommand{\D}{\displaystyle}
\newcommand{\eq}[1]{Eq.\,(\ref{#1})}
\newcommand{\fig}[1]{Fig.~\ref{#1}}
\newcommand{\km}{\mbox{{\rm\thinspace km}}}
\newcommand{\kms}{\mbox{{\rm\thinspace km\thinspace s$^{-1}$}}}
\newcommand{\ms}{\mbox{{\rm\thinspace m\thinspace s$^{-1}$}}}
\newcommand{\K}{\mbox{{\thinspace\rm K}}}
\newcommand{\tab}[1]{Table~\ref{#1}}

\newcommand{\ben}{\begin{enumerate}}
\newcommand{\een}{\end{enumerate}}
\newcommand{\bfig}{\begin{figure}}
\newcommand{\efig}{\end{figure}}
\newcommand{\beq}{\begin{equation}}
\newcommand{\eeq}{\end{equation}}
\newcommand{\mbf}{\mathbf}
\newcommand{\vare}{\varepsilon}
\newcommand{\mcal}{\mathcal}
\newcommand{\ep}{\epsilon}
\newcommand{\cs}{\mathcal{S}}
\newcommand{\csv}{\mathcal{S}_{\mathcal{V}}}
\newcommand{\cv}{\mathcal{V}}
\renewcommand{\thefootnote}{\dag}

\title{A persistent quiet-Sun small-scale tornado}
\subtitle{III. Waves}

\author{K.\,Tziotziou\inst{1}
        \and G.\,Tsiropoula\inst{1} \and I.\,Kontogiannis\inst{2}}

\offprints{K.\,Tziotziou,\\
\email{kostas@noa.gr}}

\institute{Institute for Astronomy, Astrophysics, Space Applications and Remote Sensing,
National Observatory of Athens, GR-15236 Penteli, Greece \and Leibniz-Institut f\"{u}r Astrophysik Potsdam (AIP), An der Sternwarte 16, 14482 Potsdam, Germany}

\date{Received  / Accepted }

\titlerunning{Waves in a persistent quiet-Sun tornado}

\authorrunning{Tziotziou \etal}

\abstract{Vortex flows can foster a variety of wave modes. A recent oscillatory analysis of a persistent 1.7~h vortex flow with a significant substructure has suggested the existence of various types of waves within it.}
{We investigate the nature and characteristics of waves within this quiet-Sun vortex flow, over the course of an uninterrupted 48-min observing time interval, in order to better understand its physics and dynamics.}
{We used a cross-wavelet spectral analysis between pairs of \ha\ and \ca intensity time series at different wavelengths and, hence, atmospheric heights, acquired with the CRisp Imaging SpectroPolarimeter (CRISP) at the Swedish Solar Telescope (SST), as well as the derived \ha\ Doppler velocity and full width at half maximum (FWHM) time series. We constructed halftone frequency-phase difference plots and investigated the existence and propagation characteristics of different wave modes.}
{Our analysis suggests the existence of Alfv\'{e}nic type waves within the vortex flow that propagate upwards with phase speeds of $\sim$20--30 \kms. The dominant wave mode seems to be the fast kink wave mode, however, our analysis also suggests the existence of localised Alfv\'{e}nic torsional waves, which are related to the dynamics of individual chromospheric swirls that characterise the substructure of the vortex flow. The \ha\ V-I phase difference analysis seems to imply the existence of a standing wave pattern that is possibly arising from the interference of upwards propagating kink waves with downwards propagating ones that are reflected at the transition region or the corona. Moreover, the results provide further evidence that the central chromospheric swirl drives the dynamics of the vortex flow.}
{This is the first exhaustive phase difference analysis within a vortex flow that explores the nature and dynamics of different wave modes within it. The question, however, of whether, and how, the dissipation of the derived wave modes occurs remains open, and given that such structures are ubiquitous on the solar surface, it's also important to investigate whether they might ultimately play a significant role in the energy budget of the upper layers of the solar atmosphere.}


\maketitle

\section{Introduction}
\label{intro}

Vortex flows on various, relatively small spatial and temporal scales are ubiquitous in the solar atmosphere of the quiet Sun, having been generated by the turbulent dynamics of solar convection. They are concentrated at the intergranular lanes and formed as a result of angular momentum conservation, which has as a direct consequence on the rapid rotation of the plasma sinking towards the solar interior \citep{Wede:2014}. Such flows had already been predicted in theory \citep{Sten:1975}, revealed through simulations \citep{Nord:1985,Stein:2000a, Stein:2000b},
and also reported in observations \citep{Brandt:1988, Bonet:2008}. Over the past decade, interest in quiet-Sun vortical flows has been revived thanks to their being widely observed in several photospheric and chromospheric lines, as well as the transition region (TR) and the low corona ultraviolet (UV) and extreme ultraviolet (EUV) channels \citep[e.g.][]{Bonet:2010, Attie:2009, varg:2011, Wede:2012, Park:2016, Tzio:2018}, while sophisticated numerical simulations have given important results \citep[e.g.][]{Moll:2011b, Moll:2011, Shel:2011, Kitia:2012, Wede:2009, Wede:2012}. We note that vortical coherent structures have also been  detected in photospheric turbulent flows in plage regions \citep{Chian:2014}. Quiet-Sun vortical flows have been found to be associated with motions of magnetic field concentrations referred to as bright points
\citep{Bonet:2008} and, thus, they are considered to be immediately associated with magnetic fields.
Vortex flows force the footpoints of the magnetic fields to co-rotate \citep{Wede:2014}, while the rotation is mediated from the photosphere to the low corona and creates magnetic swirling structures observed at different solar layers
\citep{Wede:2012, Tzio:2018}.
We note, however, that non-magnetic vortices with significantly different nature and dynamics than the magnetic ones, have also been reported in the literature \cite[e.g.][]{Stein:1998,Kitia:2012b}.
Nonetheless, vortex motions are important candidates for the transfer of mass, momentum, and energy, often by waves, from the subsurface to the upper atmospheric layers of the Sun. Magnetohydrodynamic (MHD) waves are considered to be a potential mechanism for the energy transport and for the heating of the solar atmosphere.

Magneto-acoustic wave propagation can be triggered by the photospheric horizontal and vertical footpoint motions of localised flux tubes \citep{Fedun:2011c}.
Analytical works, as well as simulations, have indicated that vortex structures, similarly, can foster a variety of wave modes; both periodic motions at their footpoints and the dynamics of the associated magnetic field do generate shocks and naturally drive different types of MHD waves. The generation of sausage waves has been analytically investigated both in incompressible and compressible magnetically twisted flux tubes by \cite{Erde:2006,Erde:2007}. \cite{Fedun:2011} used a high-frequency vortex motion as a driver at the footpoint of an open flux tube, with an analytically prescribed magnetic field structure, to simulate the excitation of different types of MHD wave modes, such as sausage, kink and torsional Alfv\'{e}n waves, within it. In a subsequent work, \cite{Fedun:2011b} demonstrated that the plasma structure within a vortex-driven magnetic flux tube can act as a spatial frequency filter for torsional Alfv\'{e}n waves. Such torsional Alfv\'{e}nic perturbations, as previously shown by \cite{Jess:2009}, are responsible for the non-thermal broadening of the \ha\ line profile, usually observed as variations of its full-width at half-maximum (FWHM). \cite{Vige:2012}, using a torsional driver, studied the effect of vortex-like motion in a flux tube and reported the generation of slow and fast magnetoacoustic modes. \cite{Shel:2013} using MHD simulations, including radiative transport and a non-ideal equation of state, identified horizontal motions in magnetic vortices as torsional Alfv\'{e}nic perturbations. They also steered some discussion within the community as they suggested, based on the use of test particles, that photospheric magnetic field concentrations do not produce magnetic tornadoes or a bathtub effect; this was later numerically challenged by \cite{Wede:2014}. Finally, several photospheric drivers, including Archimedean and logarithmic velocity spirals, were used by \cite{mum:2015} and \cite{mumerd:2015} to investigate the generation of Alfv\'{e}n, torsional Alfv\'{e}n waves, and other MHD waves, such as slow kink and slow or fast sausage modes, in magnetic flux tubes.

Very few works exist in the literature concerning the observational signatures of waves in vortex flows. \cite{Jess:2009} established the existence of torsional Alfv\'{e}n waves on the magnetic structure around a magnetic bright point by analysing \ha\ FWHM oscillations. \cite{Mort:2013}, using high-resolution multi-wavelength ROSA observations from the Dunn Solar Telescope at Sacramento Peak (USA), provided observational evidence (complemented by relevant numerical simulations) for chromospheric torsional Alfv\'{e}n and kink waves excited by the vortex motions of a strong photospheric magnetic flux concentration.
Recently, a detailed oscillatory analysis by \cite{Tzio:2019} of a long-lived complex vortex flow suggested the presence of magneto-acoustic waves (see below). A similar oscillatory analysis of chromospheric swirls by \cite{Shet:2019} did not provide conclusive evidence that the observed oscillations were either the response to swirling photospheric motions or the result of propagating Alfv\'{e}nic waves. \cite{Liu:2019nat}, using observations carried out with the Hinode/SOT and Swedish Solar Telescope \citep[SST;][]{Scharmer:2003a}, provided  observational evidence for the excitation of ubiquitous Alfv\'{e}n pulses by prevalent intensity swirls in the solar photosphere that propagate upwards and reach chromospheric layers.
\cite{Liu:2019AA} further supported the excitation of Alfv\'{e}n pulses by the co-spatial and co-temporal rotation of photospheric velocity swirls and magnetic swirls found in numerical simulations of the solar photosphere with the radiative MHD code Bifrost \citep{Gudi:2011}.

\begin{table*}
\caption{Summary of the used CRISP observations for the present analysis. We note that as in Paper II, no UV, EUV Atmospheric Imaging Assembly \citep[AIA;][]{Lemen:2012} and Helioseismic and Magnetic Imager \citep[HMI;][]{Scher:2012} magnetic field observations are used due to their low spatial and temporal resolution compared to CRISP observations.}
\label{table1}
\centering
\begin{tabular}{l c c}
\hline\hline
            & \ha\ 6562.81$\,${\AA}  & \ca \\
\hline
Date of observations & \multicolumn{2}{c}{June 7, 2014} \\
Observing intervals & \multicolumn{2}{c}{07:32 UT -- 08:21 UT and 08:28 UT -- 09:16 UT\tablefootmark{a}} \\
Wavelengths & \ha\ LC, $\pm0.26\,${\AA}, $\pm0.77\,${\AA}, $\pm1.03\,${\AA} &  \ion{Ca}{ii} LC, $\pm$0.055$\,${\AA}, $\pm$0.11$\,${\AA}, $\pm$0.495$\,${\AA} \\
Temporal cadence & 4~s & 4~s \\
Pixel size &  0.059$\arcsec$  & 0.0576$\arcsec$ \\
\hline
\end{tabular}
\tablefoot{
\tablefoottext{a}{There is only a small temporal offset of $\sim$2~s between the \ion{Ca}{ii} and \ha\ time series as the two lines were observed sequentially. Only the second 48-min observing interval has been used in the present analysis.}
}
\end{table*}

\begin{figure*}
\centerline{\includegraphics[width=0.9\textwidth]{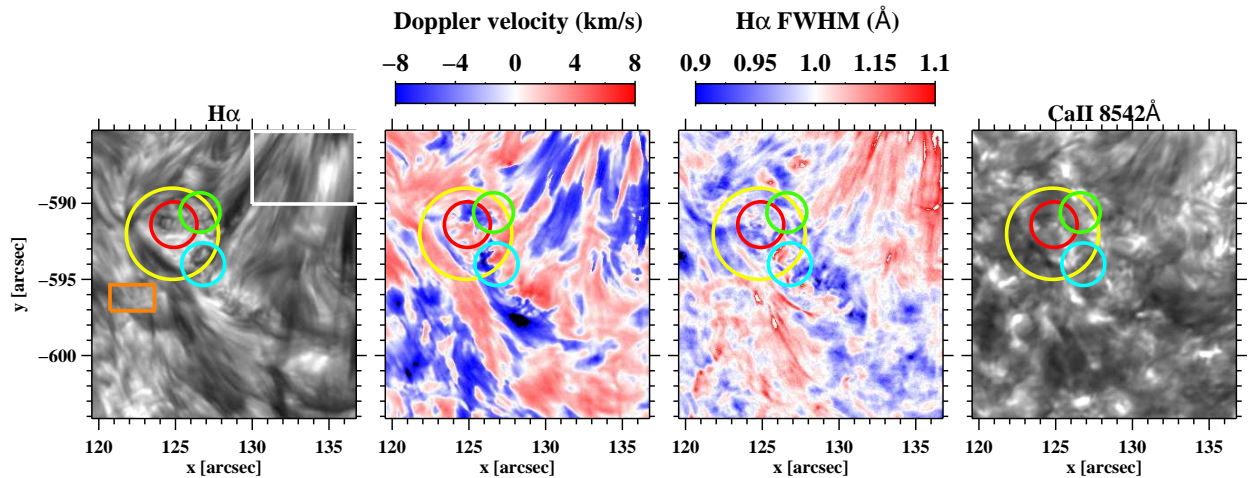}}
\caption{Snapshot of the ROI in \ha\ LC (first panel), \ha\ Doppler velocity (second panel), \ha\ FWHM (third panel), and the \ca LC. In both intensity images, black indicates structures in absorption while the grayscaling in each panel is within $\pm3\sigma$ from the respective mean value of the whole image with $\sigma$ corresponding to the respective standard deviation. The overplotted yellow circle indicates the location of the analysed conspicuous vortex flow, while smaller red, green, and cyan circles denote the approximate location of smaller swirls (substructure), extensively discussed in Papers I and II. The orange and white rectangles in the \ha\ LC panel show, respectively, the selected QSR and FSR (see text).} \label{fig1}
\end{figure*}

Several techniques have been developed for the observational study of waves on the Sun since the discovery of the 5-min acoustic oscillations by \cite{Leig:1962}. One  widely used technique for waves
that is not, however, very useful in small fields of view (FOVs) or complex structures, is $k-\omega$ diagrams, which show the relationship between the wavenumber and temporal frequency \citep[e.g.][]{Ulr:1970,Deub:1975,Deub:1979}. With higher, mainly temporal, resolution observations having become available over the years, the phase difference analysis between signals (e.g. timeseries of intensity, velocity, and, more recently, magnetic field) forming at the same or different atmospheric layers have become an important tool for the study of waves \citep[e.g.][]{mein:1976,mein:1977}. Phase differences are usually acquired through a spectral Fourier or Wavelet \citep{Torr:1998} analysis. These techniques, combined with theoretical or observational knowledge of cut-off frequencies of different waves (the lowest frequency that a wave mode can propagate) and respective formation height differences
(where necessary) provide useful information about the existence of different
wave modes and their propagation characteristics.

In two recent papers, using a multi-wavelength analysis, we meticulously investigated the characteristics and dynamics of a small-scale, 1.7-h persistent vortex flow, observed for the first time in the \ha\ line centre (LC) \citep[Paper~I;][]{Tzio:2018}, and its oscillatory behaviour \citep[Paper~II;][]{Tzio:2019}.
Our analysis revealed a rigidly or quasi-rigidly rotating, funnel-like expanding vortex flow that exhibits signatures from low chromospheric heights up to the low corona and comprises at least three recurring, intermittent smaller chromospheric swirls with typical sizes and durations. Derived oscillations, mainly in the range of 3 to 5~min peaking at $\sim$4~min and extending even up to 10~min at all heights, correspond to the cumulative action of different components such as swaying motions (with periods in the range of 200--220~s), rotation (with periods of $\sim$270~s for \ha\ and $\sim$215 for \fca), and waves. The presence of waves, mainly different from acoustic ones, such as magnetoacoustic (e.g. kink) or Alfv\'{e}n waves, have been suggested from the power behaviour within the vortex flow, compared to a reference quiet-Sun region, as a function of period and height.

In this, the third paper of the series, we use a phase difference analysis to further explore the findings of our previous papers, and mainly of Paper~II.
Our aim is to provide clues about the wave modes related to the vortex flow and of their propagation characteristics.

\section{Observations}
\label{obs}

We used the same high-cadence and high-spatial resolution SST CRisp Imaging SpectroPolarimeter \citep[CRISP;][]{Scharmer:2008} \ha\ and \ca datasets, as in Papers~I and II (see Table~\ref{table1} for a summary of the characteristics of the observations).
We note, however, that we restricted our analysis during the second observing interval (08:28 UT -- 09:16 UT) when the vortex flow is more clearly observed (see Paper~I). We also used Doppler velocities and the FWHM of the \ha\ line profile in our analysis, obtained from the \ha\ profiles according to the methodology presented in the previous two papers, where further details concerning the observations and the datacube processing and co-alignment can be found.

Figure~\ref{fig1} shows snapshots of the 17.3$\arcsec\times\,$19$\arcsec$ region of interest (ROI). The ROI is located almost at the centre of the acquired 60$\arcsec\times\,$60$\arcsec$ CRISP FOV that is centered at S38W10 on the solar disk. It contains the analysed vortex flow (marked with a yellow circle in \fig{fig1}) with an approximate center at (x,$\,$y) $=$ (124.8$\arcsec$,$\,-$592$\arcsec$) and a radius of $\sim$3\arcsec.
The `quiet' Sun region (QSR) and the region with fibril-like structures (FSR) used as comparison regions are also clearly marked in \fig{fig1}. We note that the QSR, given that the whole ROI is quite dynamic, has been set to be void of large fibril-like structures or small vortex flows, thus, approaching quiet-Sun conditions as closely as possible.

\section{Formation heights of the investigated line profiles}
\label{formh}

The formation heights of the two investigated line profiles are particularly pertinent to our analysis.
As discussed in Paper~I, the \ha\ LC has a wide formation height range centered slightly above $\sim$1.5~Mm \citep[][see their Fig.~4]{Leen:2012} with
some contributions also coming from lower heights close to 0.4~Mm. The  formation of \ha$\pm$0.26$\,${\AA}  mainly takes place somewhere between the LC formation height and the \ha$\pm$0.35$\,${\AA}
formation height at $\sim$1.2~Mm \citep[][see their Fig.~6]{Leen:2012} but it may also have some contributions from heights similar to the formation height of the photospheric \ha$\pm$0.77$\,${\AA} wavelength. Finally, \ha$\pm$1.03$\,${\AA} is photospheric as granulation is clearly visible at these wing wavelengths.

Concerning the \ca line, its LC is formed at a narrower height range at $\sim$1--1.1 Mm above the photosphere with the near-LC wavelengths of \ion{Ca}{ii}$\pm0.055\,${\AA} and \ion{Ca}{ii}$\pm0.11\,${\AA} forming within a couple of hundred kilometers below \citep[][see their Fig.~4]{Leen:2009}. This LC formation height is quite similar to the values derived by \cite{Mein:1980}. \ion{Ca}{ii} at $\pm0.495\,${\AA} is formed far lower, at low to mid photospheric heights as reversed granulation is clearly visible at these wavelengths, somewhere between 200 and 500~km according to Fig.~4 of \cite{Leen:2009} and the contribution function computed by \cite{Cauz:2008} and shown in their Fig.5. We note that \cite{Rutt:2004} report heights where reversed granulation occurs, between 250--400~km while simulations by \cite{Cheu:2007} indicate an even lower height of 130--140~km.

\section{Methodology}
\label{meth}

\subsection{Phase differences, coherency, and halftone images}
\label{cross}

The presence, characteristics, and propagation of waves are examined through a cross-wavelet analysis \citep{Torr:1998} between different pairs of signals (e.g. intensities at different heights or intensity-Doppler velocity) on every pixel of the ROI that provides the cross-power, coherence, and, eventually, phase difference as functions of time and period (frequency).

For any investigated pair $X$ - $Y$ of time series with respective wavelet transforms $W^X_n$ and $W^X_n$
the cross-wavelet transform is defined as $W^{XY}_n = W^X_n W^{Y*}_n$, where $W^{Y*}_n$ is the
complex conjugate of $W^{Y}_n$, the
cross-wavelet power as $|W^{XY}_n|$ and the phase
difference as $\phi_n= \tan^{-1}[\Im\{W^{XY}_n\}/\Re\{W^{XY}_n\}]$, where $\Re$ and
$\Im$ are, respectively, the real and the  imaginary components of the
transform. Although a Fast Fourier Transform could also provide analogous phase differences to those found by the wavelet analysis \citep{bloom:2004}, the latter has been chosen as a more suitable method due to the intermittent nature of solar oscillations. The global phase difference corresponds to the time average of the phase difference over the whole time range with cross-power weighting, as introduced by \cite{lites:1979}. This guarantees that wraparound errors are avoided when angle values just above $\pi$ are transformed into values just below $-\pi$, leading to erroneous phase differences occurring with normal averaging of individual phase differences \citep[see also][for further details]{tzio:2005}. Hereafter, when we discuss about phase differences, we refer to these global phase differences as a function of period (frequency). Finally, coherence provides a measure of the cross-correlation between the two signals and takes its values from zero (indicating no correlation) to unity.

However, because random noise always produces a non-zero coherence, a coherence threshold must be determined above
which the respective phase difference is considered reliable. For this reason, we adopt the `floor exceedance coherency approach'
described in \cite{bloom:2004} that results in a frequency-dependent coherence threshold of around 0.6 in most cases. Phase differences corresponding to lower coherence values have been discarded while halftone images, showing the distribution of the cross-power as a function of frequency and phase difference, have been constructed for the remaining, reliable phase differences, following the methodology of \cite{lites:1979}. To construct these images, the cross-power at every pixel of the considered area is summed over bins that are 3\degr\    wide for the phase difference for each frequency element and normalised to unity (using the maximum derived cross-power). Such halftone images \citep[see also,][]{konto:2016} do show the trend of phase difference as a function of frequency, while the corresponding width of the distribution provides a measure of the scatter of the cross-power \citep{lites:1993}.

\subsection{Phase speeds and cut-off frequencies}
\label{phvel}

For any frequency $f$ phase lags $\phi$ can be easily converted to time lags $\tau$ as
\begin{equation}
\tau = \frac{\phi}{2 \pi f} \,\, .
\label{tlag}
\end{equation}
In the case of vertically propagating waves between two heights with a known separation, ${\rm d}h,$ resulting to a phase difference $\Delta\phi$, time lags can in turn be related to their phase speed, $\upsilon_{\rm ph}$. The function, however, between $\Delta\phi$, $f$ and ${\rm d}h$ strongly depends on the atmospheric model used \citep[see e.g.][their Equations (1) to (4)]{Jafar:2017}. In the simplest case that concerns adiabatic propagation in an isothermal non-stratified atmosphere, the phase speed, $\upsilon_{\rm ph},$ is equal to
\begin{equation}
\upsilon_{\rm ph} = 2\pi \frac{{\rm d}h}{\Delta\phi} f \,\, .
\label{dfu}
\end{equation}
Small and constant phase differences are an indication of standing or evanescent waves. We note that a monotonic linear increase of phase differences $\Delta\phi$ as a function of frequency $f$ for a range of frequencies higher than the cut-off frequency (see below), according to \eq{dfu}, indicates upwards vertical wave propagation with a constant phase speed for this frequency range. However, when such phase differences are derived from photospheric lines, even when the physical conditions of an isothermal atmosphere exist, it has been suggested that the phase difference distribution is not always monotonically increasing with frequency \citep{mein:1976,Schm:1979,Fleck:1989}.

The propagation of waves is possible above a lowest frequency, known as the cut-off frequency, $f_{\rm 0}$. This frequency depends on the wave mode (some modes do not have a cut-off frequency) and on the ambient atmospheric conditions (e.g. temperature, density, magnetic fields). Below $f_{\rm 0}$, the expected phase difference for the associated wave mode is equal to zero as its propagation is inhibited. For acoustic waves the cut-off frequency, $f_{\rm 0}$ is 5.2~mHz and is modified as $\sim f_{\rm 0} \cos\theta$ \citep{Mich:1973,Suem:1990} in the presence of strong, inclined magnetic fields by an angle $\theta$ with respect to the solar normal (direction of gravity). In the absence of strong fields, the lowering of the cut-off frequency could also result from significant radiative damping in an isothermal atmosphere \citep[e.g.][]{Worr:2002}. On the other hand, transverse (e.g. kink and Alfv\'{e}n waves) and longitudinal magnetoacoustic waves (e.g. sausage waves)
can have a much lower cut-off frequency than the acoustic one \citep[e.g.][]{Spruit:1981, kalk:1997,McAteer:2002,McAteer:2003,Jess:2009,Stanga:2015}. Finally, torsional waves
can propagate for any period \citep{kalk:1997}. Gravity waves \citep{Miha:1981} that also have a lower cut-off frequency than acoustic ones (related to the Brunt-V\"{a}is\"{a}l\"{a} frequency) are found in much lower atmospheric heights than those investigated in the present analysis
\citep{krij:2001,Seve:2003,Rutt:2003}.

\begin{figure*}
\centerline{\resizebox{0.88\hsize}{!}{\includegraphics{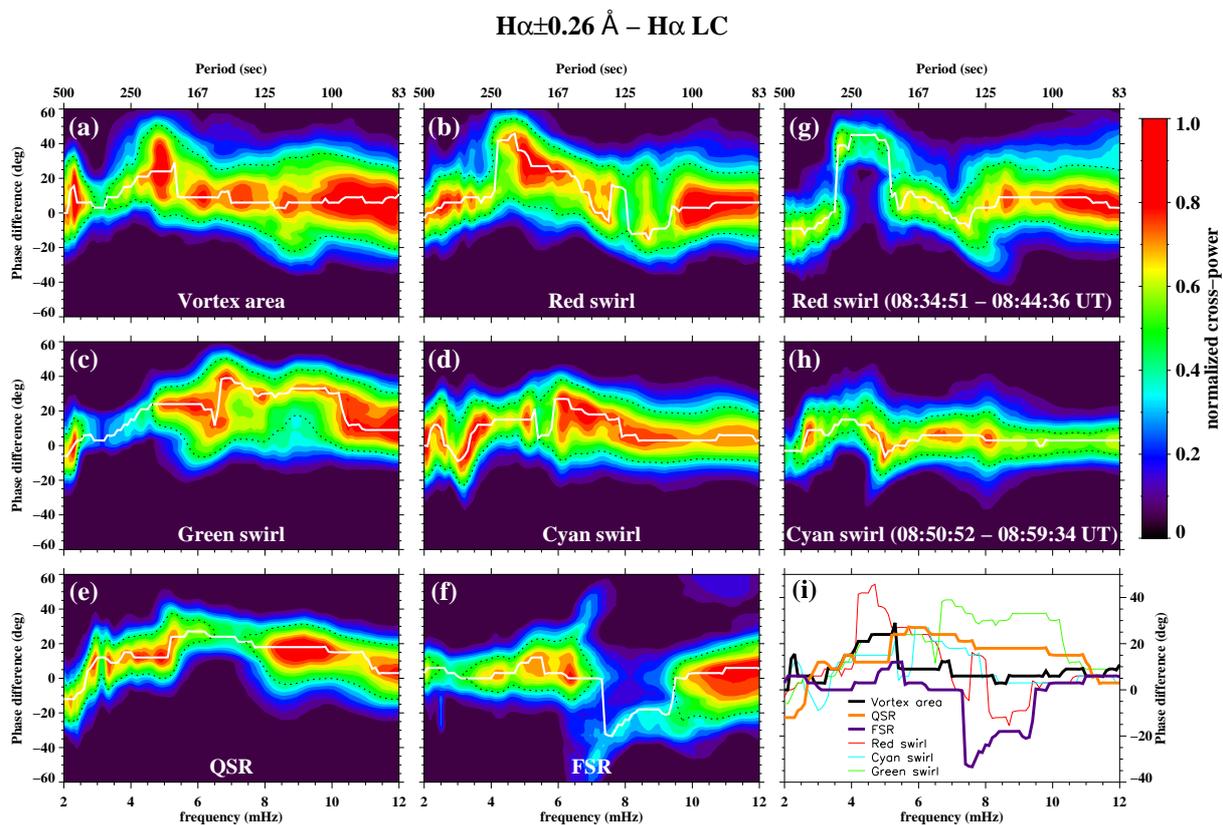}}}
\caption{Phase differences between the \ha$\pm$0.26$\,${\AA} and \ha\ LC pair within a) the vortex flow area, the areas of swirls within the red, cyan, and green circles and the QSR and FSR areas depicted in \fig{fig1} during the second observing time interval, i.e. 08:28 UT -- 09:16 UT ({\em panels a to f}) and b) for the two swirls within the red and cyan circles for specific time intervals (denoted in the respective panels) when these swirls are clearly visible ({\em panels g and h}). Both wavelengths are mainly of chromospheric origin and their formation heights are separated by a few hundred kilometers (see Sect.~\ref{formh}). Filled contours represent the cross-power distribution, while white lines denote the position of maximum cross-power, normalised to unity for each frequency element as shown by the vertical bar. The black dashed contour indicates the 50\% level from maximum cross-power, and provides a measure of the scatter as it represents the FWHM of the cross-power distribution at each frequency. Cross-power below 10\% has been disregarded while all curves have been smoothed with a running average of seven frequency points to remove spurious spikes. The corresponding phase difference of the normalised-to-unity maximum-cross power as a function of frequency, within all considered areas and for the entire considered observing interval only, is also shown in a separate panel ({\em panel i}) for an easier and direct comparison. }\label{fig2}
\end{figure*}

\section{Phase difference analysis}
\label{phdiff}

As the ROI is centered at S38W10, the line-of sight (LOS) does not coincide with the vertical to the solar surface. The estimated horizontal pixel offset from the normal to the solar surface is $\sim$0.1\arcsec per 100 km height difference which is equal to the spatial resolution of the present observations. Hence, projection effects can influence the acquired results and lead to significant errors for signal pairs involving large height separations.
Furthermore, the analysis in Paper~I indicated that the dynamics within the vortex flow changes considerably as a function of height. The vortex flow and its constituting swirls are clearly visible only in LC and near-LC chromospheric wavelengths in both investigated lines, while for wing wavelengths only some short temporal hints exist for the swirl within the red circle at \ha-0.77\,{\AA} (see Fig.4v,w of Paper~I) and \ion{Ca}{ii}-0.495\,{\AA} (see Fig.5f of Paper~I). Therefore, for all these reasons, we have restricted our analysis at phase differences between the \ha$\pm$0.26\,{\AA} -- \ha\ LC, and the \ion{Ca}{ii}$\pm$0.11$\,${\AA} -- \ca LC intensity (I--I) pairs.
We note that as the near-LC wavelengths of \ion{Ca}{ii} at $\pm0.055\,${\AA} and \ion{Ca}{ii}$\pm0.11\,${\AA} are both formed very close to the LC, we opted for the latter ones as they give the largest possible height separation.
Moreover, in order to minimise the effect of Doppler velocities on intensities, taken at a wavelength distance $\Delta\lambda$ from the LC on either wing of the \ha\ or \ca profiles, we use (similarly to Paper~II) their wavelength average (e.g. \ha$\pm$0.26 \AA\ is the average intensity at wavelengths -0.26 \AA\ and +0.26 \AA).
Both selected I-I pairs involve close wavelengths of the investigated line profiles and, hence, atmospheric heights, thus, minimising the influence of projection effects and taking into account the complex vortex dynamics.

In addition to the above intensity pairs, we also examine
intensity and Doppler velocity (I--V) and velocity and FWHM (V--FWHM) pairs, but only for the \ha\ line where this analysis is applicable. We note that we reversed the signs of Doppler velocity to correspond to the nomenclature used in early phase difference literature.
The I-I pairs that involve phase differences between different layers of the solar atmosphere are used to investigate waves and their propagation characteristics. On the other hand, I--V and V--FWHM pairs, provide physical properties of different wave modes as they relate to response signatures (derived from the line profiles) to processes, such as heating, compression, presence of torsional waves, etc.

Our phase difference analysis is mainly based on halftone images (described in Sect.~\ref{meth}) that are related to the vortex flow area and areas of the different individual swirls that constitute its substructure (see Paper~I). The QSR and FSR areas are also used for comparison. Such halftone images are derived for the entire second observing interval within all investigated areas. They have also been  derived or selected time intervals for the swirls within the red and cyan circles when these are clearly visible in both lines but not for the swirl within the green circle, which is mainly noticeable through its interaction with nearby fibril-like structures (see Paper~I). These selected intervals correspond to those of the time-slice images of Figs.~9 and 10 in Paper~I; as they are short ($\sim$10 min), the lowest investigated frequency of 2~mHz contains only slightly more than one wave period. However, for the frequency range of interest here (3--5~mHz, see below) and for higher frequencies, the number of periods is considered sufficient for the significance of the acquired results.

Before presenting below our results, we note that the oscillatory analysis of Paper~II has clearly indicated enhanced power and dominant frequencies of oscillations within the 3--5 mHz frequency band in the entire vortex area. In the three chromospheric swirls, there is a peak at $\sim$4 mHz in the \ha\ and at a slightly higher frequency in the \fca. Therefore, this 3--5 mHz frequency range (herefter, dominant frequency range) stands as the frequency range of interest.
We note that according to the nomenclature of Sect.~\ref{meth}, for any investigated pair $X$ - $Y$ (see titles and captions of all halftone figures hereafter), a positive phase difference means that $X$ leads $Y$ (and vice versa for negative values).

\begin{figure*}
\centerline{\resizebox{0.88\hsize}{!}{\includegraphics{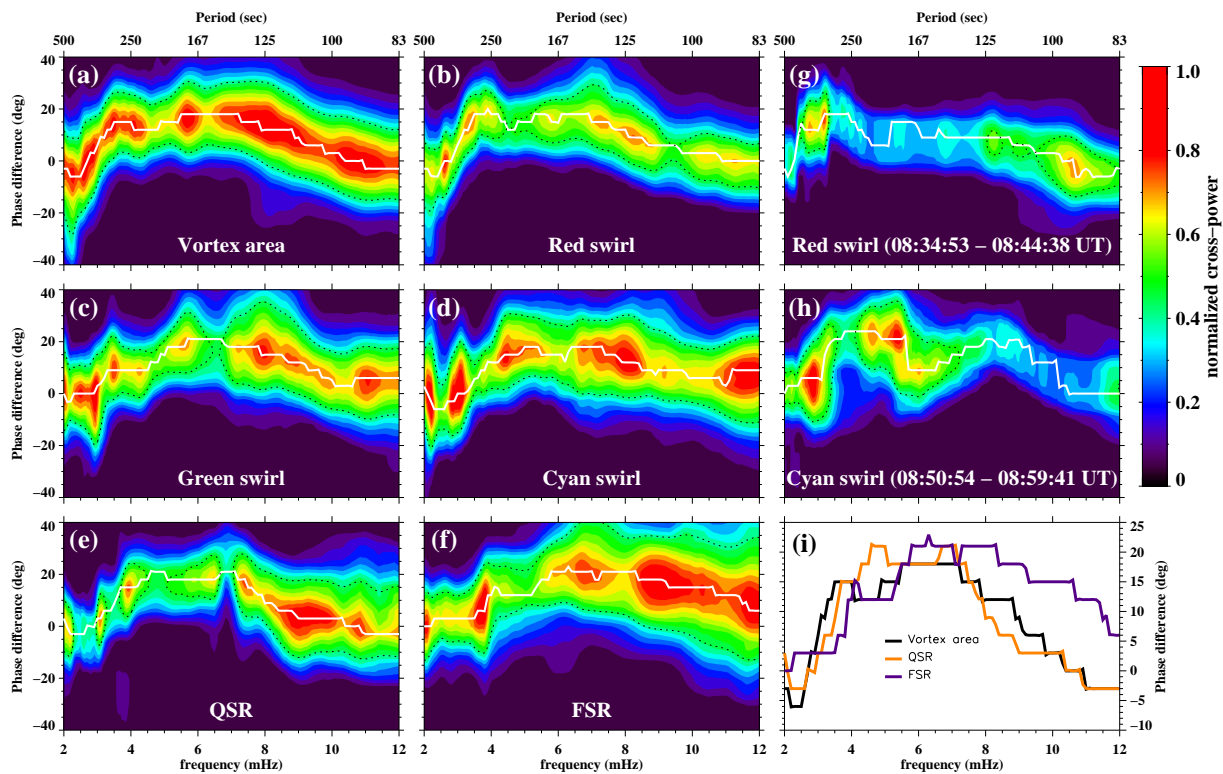}}}
\caption{Similarly to \fig{fig2} halftone images for the \ion{Ca}{ii}$\pm$0.11$\,${\AA} and the \ion{Ca}{ii} LC pair of the \ca line.
In the bottom right panel, only the corresponding curves for the vortex area, the QSR and the FSR are shown as curves for respective individual swirl areas are quite similar to these (see text).}\label{fig3}
\end{figure*}

\subsection{I--I phase difference analysis}
\label{vortbeh}

\subsubsection{Behaviour of the I--I phase differences vs frequency and inferred wave modes}
\label{phdiffbeh}

As discussed in Papers~I and II, the vortex flow comprises the swirl within the red circle and, only partly, the swirls within the cyan and green circles (see also \fig{fig1}). Parts of the vortex are sometimes free of swirling-related structures and, occasionally, its dynamics is further complicated by the presence of nearby flows from fibril-like structures. Figures~\ref{fig2} and \ref{fig3} show halftone images for the \ha$\pm$0.26\,{\AA} -- \ha\ LC and the \ion{Ca}{ii}$\pm$0.11$\,${\AA} -- \ion{Ca}{ii} LC intensity pairs, respectively, for the entire second observing time interval (panels a to f), within the vortex and swirl areas (panels a to d), as well as within the QSR and FSR areas (panels e and f) to be used for comparison. Phase differences within the vortex and swirl areas may contain also significant quiet-Sun contributions due to their intermittent dynamical behaviour.  Especially in \ha, they may be occasionally further complicated by the presence of nearby flows from fibril-like structures. For these reasons, phase differences for the swirls within the red and cyan circles, in the particular time intervals when they are continuously visible in both lines, are shown for comparison in panels g and h of Figs.~\ref{fig2} and \ref{fig3}. In all halftone panels, the overall behaviour of the phase differences versus frequency is depicted by the maximum cross-power curve (white lines). These curves are also collectively presented in panel i of Figs.~\ref{fig2} and \ref{fig3} for easier comparison.

When comparing the corresponding halftone images
for the considered \ha\ I-I pair in the vortex area and the areas of individual swirls (\fig{fig2}), we can see similarities as well as some differences between them. In all three swirls and in the vortex area, there is an increase of the phase difference in the 3--5 mHz frequency range. For the green and cyan swirls as well as for the vortex area, the increase is from $\sim$0\degr\ to $\sim$20\degr. In the red swirl, there is a small increase from $\sim$0\degr\ to $\sim$10\degr\ in the 3--4 mHz frequency range and then a jump of phase difference values to $\sim$40\degr\ in the 4--5 mHz frequency range, with low cross-power, however. A comparison of the respective panels of \fig{fig2} suggests that within the dominant frequency range, it is this swirl within the red circle that seems to dominate the phase difference behaviour within the whole vortex area. Indeed, the high phase differences, up to $\sim$40\degr, around $\sim$4.5~mHz,  with significant cross-power above 0.8, are clearly attributed to this swirl. This reinforces the argument presented in our previous works that this swirl, located at the vortex center, probably serves as a `central engine' for the entire vortex flow. Differences in the behaviour of the three swirls could also relate to the location of the swirls within the cyan and green circles that is close to the periphery of the vortex flow.

Although there is an intermittent character in the appearance of the individual swirls, the described behaviour above
seems to be indicative of their dynamics.
This can be seen from \fig{fig2} (panels g and h), where the behaviour of the phase differences versus the frequency within the two considered swirls, derived for particular time intervals when they are continuously visible, is echoed in the corresponding ones derived for the entire  observing time interval considered (\fig{fig2}, panels b and d).

The behaviour of the phase difference versus frequency for the considered \ion{Ca}{ii} I--I pair is different from that of the \ha\ I--I pair. Indeed, \fig{fig3} shows that the vortex area, the three swirl areas, and the QSR behave in a similar way when the entire second observing time interval is considered. In all these areas, there is an increase of the phase differences for frequencies $\geq$2.5~mHz and up to $\sim$4.0~mHz, where they attain values of $\sim$20\degr. This is expected, to an extent, from the observational analysis of Paper~I as a) only some intermittent swirls and not the vortex flow itself are observed in the \ca line at these wavelengths (which are both formed lower than the \ha\ I--I pair); b) these swirls are smaller than the corresponding ones seen in \ha; and c) there is almost no influence from fibril-like structures, the majority of which are barely seen.
Therefore, the behaviour of the phase differences versus frequency of all areas within the vortex flow
reflect mainly the quiet-Sun behaviour with strong photospheric contribution as it is revealed from the same low cut-off frequency of $\sim$2.5~mHz and the same general behaviour with the QSR. Minor differences between the individual swirls can be attributed to their particular dynamics. When specific time intervals are considered  (\fig{fig3}, panels g and h), likewise \ha, individual swirls show a behaviour that is reflected in the corresponding panels (\fig{fig3}, panels b and d) involving the entire considered observing interval.

The minimum frequency for which wave propagation is detected (cut-off frequency) can give clues about the wave mode (see Sect.~\ref{phvel}).
In the mainly chromospheric \ha\ I--I pair (see \fig{fig2}), the QSR and FSR show a phase difference behaviour versus frequency that is indicative of acoustic propagation and consistent with previous findings \citep{Fleck:1989,Deubner:1990}. More specifically, phase differences in the QSR are small (around 10\degr) but constant, from $\sim$ 3~mHz and up to the acoustic cut-off frequency of 5.2~mHz, indicative of standing or evanescent waves. Afterwards, they sharply increase as one would expect for acoustic wave propagation in the chromosphere. In the FSR, they are equal to zero up to $\sim$4.0~mHz suggesting that there is no wave propagation up to this frequency. Afterwards, there is a small but apparent increase of the phase difference due to the lowering of the cut-off frequency. This lowering is due to the presence of inclined magnetic fields related to the fibrilar structures that form the magnetic canopy and allow the propagation of magneto-acoustic waves for frequencies below the acoustic cut-off frequency \citep{Konto:2010b,konto:2014}. On the other hand, the behaviour of the phase differences versus frequency within the vortex area and the areas of individual swirls is different from that of the QSR and FSR. The most important result that comes from these images is that the cut-off frequency in the vortices is $\sim$3.0~mHz, which is well below the chromospheric acoustic cut-off frequency of $\sim$5.2~mHz and also well below the cut-off frequency of $\sim$4.0~mHz obtained in the FSR.

Propagation for frequencies lower than the acoustic cut-off frequency of 5.2~mHz, as detailed in Sect.~\ref{phvel}, is an indication of the existence of either non-acoustic wave modes or magneto-acoustic waves that are allowed to propagate due to the lowering of the acoustic cut-off frequency in the presence of inclined magnetic fields or an atmosphere with significant radiative damping. The significantly lower cut-off frequency within the vortex flow and the individual swirls points directly to the existence of non-acoustic wave modes
and, more specifically, to Alfv\'{e}nic type modes.
This confirms the results of the oscillatory analysis of Paper~II that have clearly suggested the presence of mainly transverse wave modes that are different than the acoustic ones without, however, completely ruling out the presence of some acoustic power. The natural interpretation of the present results, when combined with the swaying motions observed in the area (see e.g. Fig.~9 in Paper~I and Fig.~3 in Paper II), is that the areas experience kink MHD oscillations. Kink waves are the only wave modes that can displace the axis of the three individual swirls and of the vortex flow as a whole and have such a lower cut-off frequency. As stated in Sect.~\ref{intro}, the ubiquitous vortex flows can effectively excite kink waves that propagate upwards. The above results provide clear evidence that vortical motions generate kink waves that propagate
from the lower layers to the chromosphere.

\subsubsection{Wave propagation characteristics}
\label{upwave}

Despite some obvious differences (discussed in the previous section) in the behaviour of the I--I phase differences versus frequency between the vortex flow and the individual swirls, as well as in the different lines and wavelengths, phase differences, when they are not zero and notwithstanding the significant scatter, are mainly positive with significant cross-power.
These prevalent positive phase differences suggest upwards propagation as the lower formed intensity variations lead the higher formed ones. Moreover, linear increases of phase difference with frequency indicate upwards propagation with a constant phase speed.

Indeed, in \ha\ (\fig{fig2}) and within the dominant frequency range of 3--5~mHz, we see positive phase differences up to $\sim$40\degr\ with highly variable behaviour, however, as a function of frequency within the swirl denoted with a red circle. Mainly positive phase differences but with lower values ($\sim$10\degr-- 20\degr) are also observed within the swirls denoted with the green and cyan circles despite some negative phase differences for the latter for frequencies lower than 3.5~mHz. As a consequence, positive phase differences are also seen within the whole vortex area. For the \ion{Ca}{ii} I--I pair in \fig{fig3}, likewise \ha, phase differences within the vortex area and individual swirls, are again positive with significant cross-power present but with lower values than in \ha.
We note that even in the QSR and FSR, in both lines, we clearly have upwards wave propagation above the respective cut-off frequencies.

\begin{figure}
\centerline{\resizebox{0.8\hsize}{!}{\includegraphics{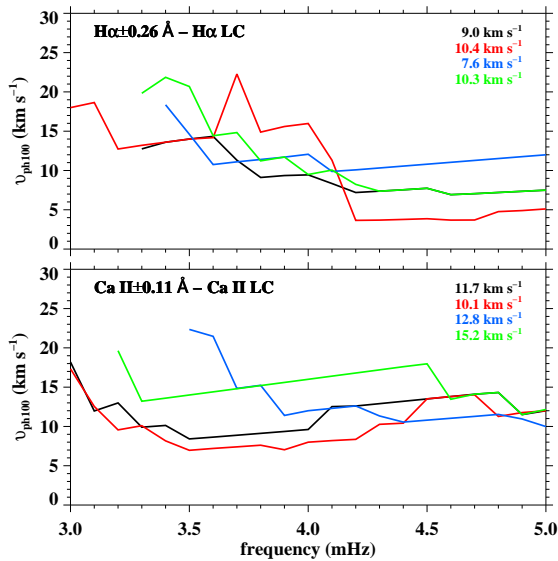}}}
\caption{Phase speeds $\upsilon_{\rm ph100}$ (in \kms) for a height difference of 100 km derived from the phase difference corresponding to the maximum cross-power of the considered I--I pairs (white lines in Figs.~\ref{fig2} and \ref{fig3}) within the dominant frequency range and for different areas. Here, $\upsilon_{\rm ph100}$ has been calculated only for phase differences higher than 10\degr\ within the vortex area (black curves), and the swirls within the red, cyan, and green circles (respective colour curves) shown in \fig{fig1}. Coloured legends provide the weighted mean $\upsilon_{\rm ph100}$ values with the respective normalised cross-power as weight. The derivation of actual speeds, $\upsilon_{\rm ph},$ further requires the actual formation height separation, ${\rm d}h,$ as they are equal to $({\rm d}h$/100~km)$\times\upsilon_{\rm ph100},$ according to \eq{dfu}.}\label{fig4}
\end{figure}

Figure~\ref{fig4} shows the derived phase speeds, $\upsilon_{\rm ph100},$ for a height difference of 100~km used as a reference (an approach chosen as formation height estimates may change in the future with better line formation codes becoming available). Actual values of $\upsilon_{\rm ph}$ are equal to $({\rm d}h$/100~km)$\times\upsilon_{\rm ph100},$ with ${\rm d}h$ representing the formation separation height of the two wavelengths of the considered I--I pair. This ${\rm d}h$, according to Sect.~\ref{formh}, is in the range of 200--300~km between \ha$\pm$0.26$\,${\AA} and \ha\ LC and $\sim$200~km between \ion{Ca}{ii}$\pm0.11\,${\AA} and \ion{Ca}{ii} LC.
Weighted-mean phase speeds do not substantially differ within the vortex and the different swirl areas. Actual phase speeds, when the above ${\rm d}h$ are considered,
derived from the lower forming \ion{Ca}{ii} I--I pair are generally similar to those derived from the \ha\ I--I pair and on the order of 20--30\kms\ or even higher at certain frequencies. Phase speeds in \ha\ tend to be slightly lower, especially within the swirl denoted with the red circle, when only frequencies above $\sim$4~mHz are considered. Then the weighted mean $\upsilon_{\rm ph100}$ becomes 7.6, 5.7, 11 and 7.7\kms\ within the vortex and the swirls within the red, cyan and green circles, respectively. These derived phase speeds of 20--30\kms\ are definitely higher than the local sound speed, which is of the order of 10\kms, but are compatible with typical values of Alfv\'{e}nic type speeds at these atmospheric heights \citep[e.g.][their Figs.~7, 3, and 2, respectively]{depo:2001,Cranmer:2005,Arber:2016}.

Uncertainties in the derivation of phase speeds could rise from the assumed physical conditions (see Sect.~\ref{phvel}), and from the simplified assumption of vertical wave propagation. Although the actual three-dimensional magnetic structure of vortex flows cannot presently be inferred from observations, simulations indicate that the magnetic field is mostly far from vertical and it is, in fact, rather twisted and possibly expanding with height. Wave propagation in such a configuration has only been treated under certain assumptions and
for specific wave modes either analytically \citep[e.g.][]{Erde:2006,Erde:2007,Chere:2018} but mainly numerically \citep[e.g.][]{Mura:2018}. Results indicate that the various MHD waves are strongly guided by the local magnetic field \cite[see e.g.][their Table 4]{Bogd:2003}. Therefore, phase differences at the same image pixel employing signals at different heights, for instance, phase differences between different wavelengths of a line profile, should always be treated and explained with extreme caution in terms of wave propagation.

\begin{figure}
\centerline{\resizebox{1.\hsize}{!}{\includegraphics{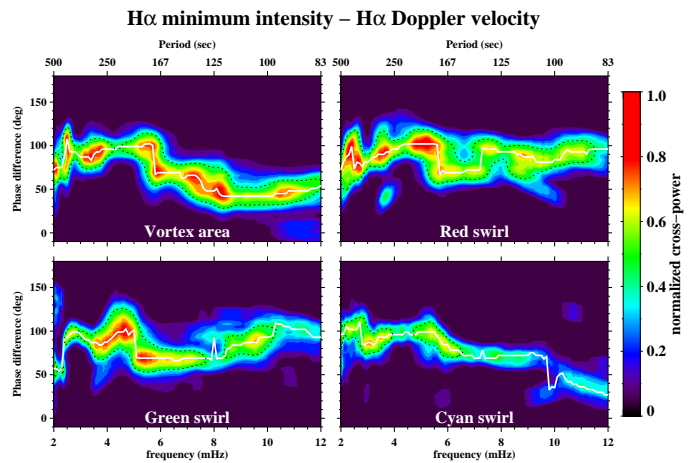}}}
\caption{Similarly to \fig{fig2} halftone images for the \ha\ minimum intensity -- \ha\ Doppler velocity pair but only for the vortex area and the individual swirl areas for the entire  observing interval considered.} \label{fig5}
\end{figure}

\subsection{I--V phase difference analysis}
\label{phdiffiv}

Phase differences between intensity and velocity (I--V) can also be used to infer the properties of waves. Generally, \ha\ intensity is considered to be a proxy of temperature and density. However, its dependence on temperature is stronger at the wings of the line and weakens towards the LC \citep{Leen:2012}, further complicating the interpretation of the I--V phase differences.

Considerable work has been done in the past on I--V phase differences for acoustic waves and, especially, for p-modes. It goes back to \cite{Whit:1958} and \citet[][their Fig.~6]{Holl:1975} who showed that for running upwards waves with periods $\sim$5 min, temperature (as a proxy of intensity) and velocity should be largely in phase and have positive phase differences. For standing waves, on the other hand, intensity should lead velocity by 90\degr\ in the adiabatic limit. In the isothermal case and when the timescale for heat losses is short, the phase difference may reach up to 180\degr\ while negative phase differences between -180\degr\ and -90\degr\ indicate downwards propagation. This phase difference value can be interpreted as a superposition of ascending and descending waves reflected at a chromospheric or transition region boundary with the dense photosphere acting as the lower boundary. \cite{mein:1976} and \cite{mein:1977}, in discussing the V-I phase behaviour of upper chromospheric lines, suggested the generation of standing waves by the transition region. \cite{mein:1977}, however, also pointed out the inherent difficulties of such a single reflecting boundary, which was later also heavily debated by \cite{Debn:1996}.  \cite{Fleck:1989} also reported the observations of acoustic standing waves in the lower solar chromosphere by measuring a 90\degr\ phase difference between the brightness and velocity oscillations of the \ion{Ca}{ii} lines. The formation of a nearly standing wave pattern in the chromosphere through the interference of upwards propagating Alfv\'{e}n waves with reflected, downwards-propagating ones has been suggested by \cite{Holl:1981}; reflection occurs at atmospheric levels, where the Alfv\'{e}n speed changes rapidly as we move from the chromosphere to the transition region. \cite{Fuji:2009}, using Hinode/SOT spectropolarimetric observations, considered fluctuations of the LOS magnetic flux and velocity in magnetic flux tubes and found phase differences of $\sim$ -90\degr\ or $\sim$ 90\degr. They conjectured that these result from the superposition of an ascending and descending kink wave. The former
is reflected at the chromospheric-coronal boundary and the ascending and descending superposed waves form standing waves at the line formation layer. Whether the phase is -90\degr\ or 90\degr\ depends on the distance of this layer from the reflecting boundary.

Figure~\ref{fig5} shows the acquired phase difference as a function of frequency between the \ha\ minimum intensity and the \ha\ LC Doppler velocity within the vortex area and the three individual swirls. Phase differences are always positive meaning that intensity always leads upwards velocity. Moreover, at least for frequencies within the dominant 3--5~mHz range, they are distributed around 90\degr.
For frequencies higher than $\sim$5~mHz the phase difference generally drops as a function of frequency, but remains always positive. In some cases, for frequencies above 7 or 8 mHz, it becomes constant, while for the swirl within the red circle it again attains values close to $\sim$90\degr.
As already discussed (Sect.~\ref{vortbeh}),
the observed fluctuations are consistent with kink wave modes. The observed I--V phase differences of 90\degr\ (\fig{fig5}),
is an indication of the existence of standing kink waves within the vortex and the individual swirl areas at the formation height of the \ha\ minimum intensity. These waves can be produced by the upwardly propagating kink waves
which, after being reflected in the transition region or corona propagate downwards and form the standing waves in the \ha\ minimum intensity formation layer.
However, standing waves due to the superposition of upwards and downwards propagating magnetoacoustic waves produced by the p-modes cannot be excluded.

Reflections of upwardly propagating magnetoacoustic waves have been examined in the case of fibril-like structures like the ones at the west and north-west border of the vortex flow (see Paper~I). Such inclined structures form the magnetic canopy and play a crucial role not only in the reflection, but also in the transmission and refraction of waves \citep{konto:2014,konto:2016}. Unfortunately, as we lack simultaneous high-spatial resolution observations of the magnetic field it is not possible to infer any possible interaction between MHD waves and the magnetic field within the vortex structure, or even with the nearby fibril-related magnetic canopy. However, we note that such an investigation would not be an easy task because of the complicated form of the magnetic field lines and the plasma motion inside vortex magnetic structures.

\begin{figure}
\centerline{\resizebox{0.8\hsize}{!}{\includegraphics{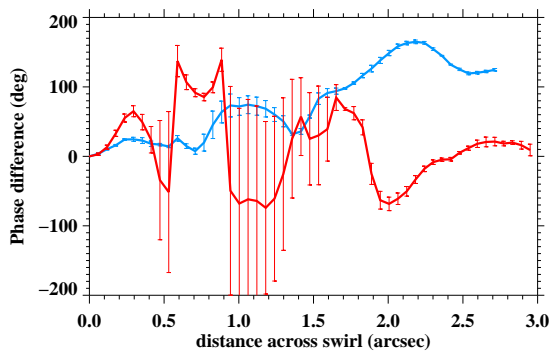}}}
\caption{Average phase difference of FWHM oscillations around the dominant frequency of 4~mHz ($\pm$0.5~mHz) as a function of distance across the diameters of the swirls within the red (red curve) and cyan circle (cyan curve). The FWHM phase differences are derived relative to each swirl's left edge that is used as a reference point, with the ends of both curves ideally associated with opposite sides of the waveguide in the case of torsional Alfv\'{e}n perturbations. Only FWHM oscillations at selected time intervals when these swirls are clearly visible (see \fig{fig2}, panels g and h) have been considered for the phase differences derivation. Overplotted error bars correspond to the respective standard deviation of the FWHM oscillations within the considered frequency range.}\label{fig6}
\end{figure}

\subsection{FWHM oscillations and phase differences between the \ha\ Doppler velocity and FWHM}
\label{fwhm_osc}

In Paper~II, we showed that the FWHM parameter derived from the \ha\ line profiles shows oscillations and enhanced power within the vortex area in the same dominant frequency range as the intensities. Variations of the non-thermal width of a line, therefore, of the FWHM, can be caused by a number of mechanisms, such as Kelvin-Helmholtz instabilities \citep{Kuri:2015,Kuri:2016}, transverse displacements, and rotational motion due to kink waves \citep{Goos:2014} or torsional Alfv\'{e}n waves \citep{Zaqa:2003,Zaqa:2007,Jess:2009}. Investigations of the connection between periodic transverse displacements and FWHM variations were mainly performed in spicules at the solar limb, where transverse displacements of their axis correspond to observed LOS Doppler velocities and are easy to measure. In such cases, the largest FWHM is produced at zero displacement from the equilibrium position when the torsional velocity is at its maximum.

On the solar disk, kink waves are manifested by the transverse displacements of the magnetic structure axis, which can be observed and measured by an imaging instrument. When the axis of the investigated structure is not parallel to the LOS, as in our case, such motions constitute a component of the derived Doppler velocity. Torsional Alfv\'{e}n waves, on the other hand, do not cause any bulk displacement of the structure about its central axis or variations in the intensity (i.e. density) but, rather, they cause purely axisymmetric twisting motions and, hence, they are not easily detected. One way to detect them is if they show both red and blue Doppler shifts simultaneously at the opposite edges of the structure, with its axis not in parallel to the LOS. Due to the spatial integration, this creates a periodic non-thermal broadening of the spectral line. In this case, the 180\degr\ phase delay of FWHM oscillations that are produced at the opposite boundaries of the waveguide that outline constant magnetic surfaces \citep{Copil:2008, Door:2008} are evidence of torsional Alfv\'{e}n waves. This has been clearly shown in Fig.~4 of \cite{Jess:2009}, where the phase differences stemming from FWHM oscillations increase from 0\degr\ to 180\degr\ across the diameter of a magnetic bright point.

\begin{figure}
\centerline{\resizebox{1.\hsize}{!}{\includegraphics{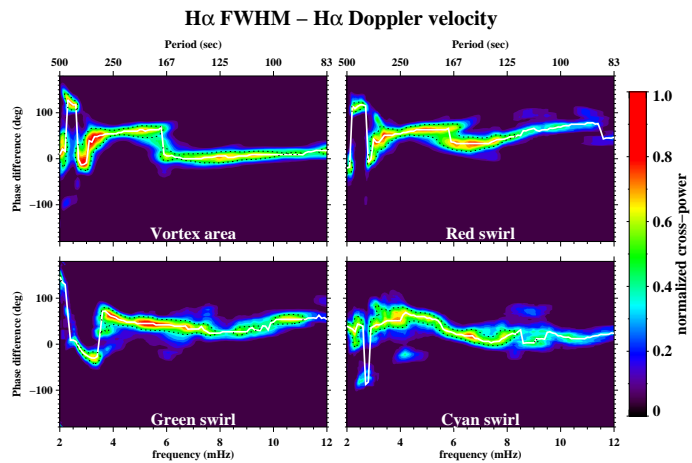}}}
\caption{Similarly to \fig{fig2} halftone images for the \ha\ FWHM -- Doppler velocity pair but only for the vortex area and the individual swirl areas for the entire   observing interval considered.} \label{fig7}
\end{figure}

Figure~\ref{fig6} shows a plot of the average phase difference of FWHM oscillations as a function of diameter for the swirls within the red and cyan circles. Only the swirl within the cyan circle shows a monotonic increase of the FWHM phase difference from one edge of the structure to the opposite edge, which nearly reaches a value of 180\degr.
This result could be suggestive of the presence of torsional Alfv\'{e}n waves within this swirl. We note that as indicated in Paper~I, the rotation of this swirl sometimes changes from counterclockwise to clockwise, which is contrary to the swirl within the red circle that shows a regular clockwise rotation and it is clearly seen only at  larger heights than the swirl within the red circle. Both of these effects could also result from the action of torsional waves. The corresponding behaviour for the swirl within the red circle in \fig{fig6} differs completely. No monotonic increase of the phase difference exists across its diameter, suggesting that there is no direct indication of torsional Alfv\'{e}n waves within this swirl,which probably acts as a central driver of the observed flow.

We note that in Paper~II, variations of the rotational period as a function of the radius were suggested within the vortex flow.
Such variations could trigger Kelvin-Helmholtz instabilities that would also manifest themselves as small-scale vortices within a large vortex flow. Although such a possibility cannot be totally ruled out, the size of the observed swirls and the compelling evidence provided above point to the existence of torsional Alfv\'{e}n waves.

Figure~\ref{fig7} shows the derived phase differences for the \ha\ FWHM -- Doppler velocity pair within the vortex area and the different swirls. The halftone images show similarities with the corresponding images shown in (\fig{fig5}).
To our knowledge, no similar observational results exist in literature for a direct comparison while a theoretical interpretation is still missing. The phase differences in all areas attain values of $\sim$ 60\degr-70\degr\ in the frequency range of 3--5~mHz. The FWHM of a line includes thermal and non-thermal broadening. Thermal broadening depends on the temperature inside the structures, while non-thermal broadening is related to turbulent motions, waves, and various inhomogeneities. As stated in Sect.~\ref{phdiffiv}, \ha\ intensity has a stronger dependence on temperature at the wings of the line, weakening towards the LC. Differences between \fig{fig5} and \fig{fig7} may result from different influences of the thermal and non-thermal components on intensity and FWHM.

\section{Discussion and conclusions}
\label{wave}

Observations and simulations (see Sect.~\ref{intro}) demonstrate that due to spiralling motions at the photosphere plasma is propagating upwards within vortical magnetic structures, using twisted magnetic field lines as guides. The complex interaction between these rotating plasma flows and vortex-related magnetic fields are thought to be the source of several MHD wave modes. Different, mainly Alfv\'{e}nic type modes, such as kink waves, sausage modes, and torsional Alfv\'{e}n waves have been widely suggested in literature to be excited by vortex flows \citep{Fedun:2011, Fedun:2011b, Shel:2013, Verth:2016}. The kink mode corresponds to a bulk motion of the plasma within and outside a magnetic structure and, amongst others, displaces the whole structure in the transverse direction. Sausage modes are typically identified by periodic fluctuations of the area of the magnetic structure. Torsional Alfv\'{e}n modes, on the other hand, can exist independently on each magnetic surface of the structure and do not displace it as a whole \citep[e.g.][]{Erde:2007b, Door:2008,Rude:2009}. Despite these behavioural differences, kink modes are considered to be of Alfv\'{e}nic type as phase speeds are similar to those of Alfv\'{e}n waves and magnetic tension is also the restoring force. These are characteristics that often lead to a confusion between the two modes in literature.

In Paper~II, the power and oscillatory analyses of a vortex area have clearly indicated the presence of significant oscillatory power in the range of 3 to 5~mHz, which peaks around 4~mHz, and reflects the cumulative action of different components, such as swaying (transverse) motions, rotation, upwards motions, and waves. Moreover, the behaviour of power within the vortex flow as a function of period and height, when compared to a reference quiet-Sun region, clearly suggested the existence of waves that are substantially different from the acoustic ones, such as magnetoacoustic (e.g. kink) or Alfv\'{e}n waves.

Observational evidence on the types of MHD wave modes that vortex-related flux tubes can support is very important. Not only because magneto-seismology allows us to obtain a diagnostic insight into the properties of the solar plasma, but also because MHD waves channeled through vortex flux tubes are considered to play a key role in the heating of the solar atmosphere as they may carry significant amounts of energy throughout the solar layers  \citep{Wede:2012,Liu:2019nat}. A practical tool for identifying wave modes from observations and their propagation characteristics is a phase difference analysis. In this work, we perform such an analysis between pairs of different parameters within the vortex area and the three individual swirls within it. More specifically, we investigate phase differences between intensity pairs (I--I) at different wavelengths and, hence, sampled layers of the solar atmosphere, in the two observed lines (i.e. \ha\ and \fca), and this analysis is also conducted in a QSR and an FSR for comparison. We also perform a phase difference analysis between pairs of \ha\ minimum intensity--Doppler velocity (I--V) and FWHM--Doppler velocity (FWHM--V) in the vortex area and the three individual swirls.

The behaviour of I--I phase differences with frequency within the vortex-related areas has different characteristics than those within the QSR and the FSR. In the QSR, there is wave propagation for frequencies above the acoustic cut-off frequency of 5.2\,mHz, while in the FSR, there is wave propagation for frequencies above a lowered cut-off frequency of $\sim$ 4\,mHz due to the role of inclined magnetic fields in the propagation of magnetoacoustic waves \citep{Suem:1990,konto:2016}. In the vortex area and the individual swirls, the present analysis clearly demonstrates and further supports the existence  of other than acoustic wave modes. All results indicate the presence of upwards propagating waves in the frequency range 2.6 - 5\,mHz, with a cut-off frequency of $\sim$ 2.6\,mHz, which is much lower than the acoustic cut-off frequency of 5.2~mHz. This lowered cut-off frequency cannot be attributed to acoustic waves travelling within inclined magnetic fields or atmospheres with excessive radiative damping. Such a low cut-off frequency is, however, consistent with the theoretical expectations for kink waves \citep{Spruit:1981}. In addition, acquired phase speeds of upwards propagating waves are mainly in the range of 20--30 \kms\ which are compatible with Alfv\'{e}n speeds at the considered solar atmospheric heights and further support the interpretation of the propagating waves as fast kink waves. It is worth remarking that such an interpretation is also in agreement with the evidence for the existence of transverse motions within the vortex flow, reported in the previous two papers (see e.g. Figs.9 to 12 of Paper~I and Fig.3 of Paper~II). These results fit the picture of a lateral displacement of the axis of the vortex structure that cannot be attributed to other types of waves, such as torsional Alfv\'{e}n or sausage waves. The existence of sausage waves can be further excluded as: a) no periodic fluctuations of the area of the vortex have been observed, with the vortex area clearly remaining, at least visually, constant (see Paper~I); and b) the time-slice images of Fig.~9 of Paper~I do not show symmetric spiral variations around the vortex central position as would be expected in this case.

Apart from the propagating fast kink waves within the vortex area and the three swirls, revealed from the I--I phase differences, the systematic phase differences between the \ha\ intensity minimum and the Doppler velocity, which are equal to $\sim$ 90\degr\ in the same examined areas, are suggestive of the existence of standing waves. These waves may result from the superposition of upwards and downwards propagating waves and are formed at the formation layer of the \ha\ minimum intensity. We suggest that the ascending waves are kink waves that after being reflected in a boundary found somewhere in the TR or in the corona move downwards with this superposition of counter propagating waves resulting in the formation of standing waves. Standing waves have been detected before in the lower solar atmosphere, for example, by  \cite{Fuji:2009} who examined phase differences between magnetic field and velocity in pores and intergranular magnetic structures and interpreted them as due to kink or sausage waves. However, standing waves due to counter propagating magnetoacoustic waves produced by the p-modes cannot by excluded by the present study.

The presence of kink waves cannot exclude the co-presence of torsional Alfv\'{e}n waves within the structure since such waves could act in individual magnetic surfaces of the structure. Unfortunately, torsional Alfv\'{e}n waves cannot be easily detected either as intensity or Doppler velocity variations because usually the spatial scale or the temporal periodic variation of the mode are not resolved.  Indirectly, the existence itself of substructure in the form of several smaller chromospheric swirls within the conspicuous vortex flow could be attributed to torsional Alfv{\'e}n waves, as explained in the work of \cite{Fedun:2011b}. Moreover, oscillations of the FWHM of the \ha\ line profile, which have been clearly seen and described in Paper~II, can be attributed to torsional Alfv{\'e}n waves \citep{Jess:2009}. Oscillations of the FWHM of the \ha\ profile are mainly caused by periodic fluctuations of non-thermal broadening, which, in turn, can be caused by different reasons, with rotation among them. The phase difference analysis between the FWHM and Doppler velocity oscillations provides a compelling argument that, at least, the swirl within the cyan circle results from the action of torsional Alfv\'{e}n waves. A phase delay of the FWHM oscillations equal to 180\degr, is found on the opposite edges of this particular swirl. Such a phase difference increase from 0\degr\ to 180\degr\ of the FWHM oscillations, found across the diameter of a magnetic element, was attributed to torsional Alfv\'{e}n waves by \cite{Jess:2009}. We note, as the simulations indicate, that the main driving force behind such Alfv\'{e}n waves could mainly be attributed to the rotation of the vortex structure rather than its transverse motion, which is attributed to kink waves.

In conclusion, the phase difference analysis performed in this work, combined with the oscillatory analysis (Paper II) and the analysis of the vortex dynamics (Paper I), suggest the coexistence of MHD waves of Alfv\'{e}nic type both propagating and standing. The dominant type of waves seems to be fast kink waves that propagate upwards and affect the oscillatory behaviour of the entire vortex structure. Localised torsional Alfv\'{e}n waves related to the dynamics of individual chromospheric swirls within the vortex structure seem to coexist. Moreover, the existence of a standing wave pattern at the height of formation of the \ha\ minimum intensity is implied by the present analysis. This pattern could arise from the superposition of upwards propagating kink waves with downwards-propagating ones after being reflected at the TR or corona.
Numerical simulations have suggested that different MHD wave modes such as kink waves and torsional Alfv\'{e}n waves may be excited in the same vortex tube and can coexist simultaneously \citep{Fedun:2011}. From observations of spicules at the limb, the coexistence of both kink and torsional Alfv{\'e}n waves \citep{depo:2012}, as well as propagating and standing transverse oscillations \citep{Okam:2011} have been reported. Both transverse propagating and standing wave modes have also been detected in on-disk chromospheric mottles \citep{Kuri:2013}.

As vortex flows act as guides for the propagation of different types of MHD waves, a possible dissipation of the energy carried by them is of great interest for estimating the contribution of vortical structures to the energy budget of the upper layers of the solar atmosphere. Our findings reveal the upwards propagation of fast kink waves with frequencies above 2.6\,mHz at the chromospheric heights considered. These waves may carry a considerable amount of energy above the heights of formation of the cores of the \ca or \ha\ lines. Our observations and analysis, however, cannot elucidate whether and where this energy is dissipated; this is certainly not a straightforward task and it is certainly out of the scope of the present study. We find that fast kink modes most probably are partially reflected in a boundary in the TR or corona.
It is not clear which amount of the waves propagating upwards reach the corona or whether they are (partially or entirely) dissipated below it and by releasing their energy play a role in heating their local surroundings. It has been shown that reflection occurring in the TR produce in the chromosphere a pattern of counter-propagating waves that are subject to nonlinear wave-wave interactions. The waves decay into turbulence which may significantly increase the heating rates in the chromosphere, while part of the wave energy is transmitted through the TR and produces turbulence in the corona \citep{Matt:1999,Ball:2011}. Torsional Alfv{\'e}n waves may also play an important role in the energy propagation through the solar atmosphere and its heating. \cite{Sole:2017}  investigated the reflection, transmission, and dissipation of torsional Alfv{\'e}n waves propagating in expanding flux tubes from the photosphere up to the low corona. They have shown that the energy propagation of the torsional waves goes through three different regimes depending on the frequency of the waves. Assuming that the waves are driven below the photosphere, they found that low frequencies are mainly reflected back to the photosphere, while intermediate frequencies are able to be efficiently transmitted to the corona, and high frequencies are completely damped in the chromosphere.

The work presented in this series of papers (Paper~I, Paper~II, and this paper) represents the first exhaustive analysis of the dynamics, oscillations, and waves of a persistent vortex flow. We note, however, that both a theoretical and analytical verification of the acquired results is still missing and continues to be necessary. Our analysis has greatly profited from the high spatial and temporal resolution observations provided by the SST. New high-resolution, multi-line observations in more wavelengths across the line profiles that include simultaneous spectropolarimetric observations are imperative to gaining a broader understanding of the dynamics of swirling structures, which seem to be ubiquitous in the solar atmosphere. Such observations could provide further clues about the generation and propagation of the various types of MHD waves within them and provide answers to the question of whether they play any significant role in the energy budget of the solar atmosphere.

\begin{acknowledgements}
The authors would like to thank the International Space
Science Institute (ISSI) in Bern, Switzerland, for the hospitality provided to the
members of the team on ``The Nature and Physics of Vortex Flows in Solar
Plasmas''. KT and GT acknowledge support of this work by the project ``PROTEAS II'' (MIS 5002515), which is implemented under the Action ``Reinforcement of the Research and Innovation Infrastructure'', funded by the Operational Programme ``Competitiveness, Entrepreneurship and Innovation'' (NSRF 2014-2020) and co-financed by Greece and the European Union (European Regional Development Fund). This research was also supported by IKYDA2020, an action program between the German Academic ExchangeService (Deutscher Akademischer Austauschdienst - DAAD) and the Greek State Scholarship Foundation (I.K.Y). IK acknowledges support by the Deutsche Forschungsgemeinschaft (DFG) grant DE~787/5-1.
The Swedish 1-m Solar Telescope is operated on the island of La Palma by the Institute for Solar Physics of Stockholm University in the Spanish Observatorio
del Roque de los Muchachos of the Instituto de Astrofisica de Canarias. Many thanks to the referee for the insightful comments and suggestions on improving the paper.
\end{acknowledgements}

\bibliographystyle{aa}
\bibliography{tziotziou} 

\end{document}